# Tunable Nanophotonic Devices and Cavities based on a Two-Dimensional Magnet

*Ahmet Kemal Demir[1†], Luca Nessi[1†], Sachin Vaidya[1], Connor A. Occhialini[1], Marin Soljačić[1], Riccardo Comin[1\*]*

1. Department of Physics, Massachusetts Institute of Technology, Cambridge, Massachusetts 02139, United States

   †These authors contributed equally to this work.
   *To whom correspondence should be addressed: rcomin@mit.edu

**Abstract**

Central to the field of nanophotonics is the ability to engineer the flow of light through nanoscale structures. These structures often have permanent working spectral ranges and optical properties that are fixed during fabrication. Quantum materials, with their correlated and intertwined degrees of freedom, offer a promising avenue for dynamically controlling photonic devices without altering their physical structure. Here, we fabricate photonic crystal slabs from CrSBr, a van der Waals antiferromagnetic semiconductor, and demonstrate unprecedented in situ control over their optical properties. Leveraging the combination of the exceptionally large refractive index of CrSBr near its excitonic resonances and its tunability via external fields, we achieve precise manipulation of photonic modes at near-visible and infrared wavelengths, showcasing a new paradigm for nanophotonic device design. The resulting guided resonances of the photonic crystal are tightly packed in the spectrum with very small mode volumes, are highly tunable via external magnetic fields, and exhibit high Q-factors exceeding 500. These resonances self-hybridize with the excitonic degrees of freedom, resulting in intrinsic strong light-matter coupling. Our findings underscore the potential of quantum materials for developing in situ tunable photonic elements and cavities.

**Introduction**

Harnessing emerging quantum materials in photonics represents a promising research frontier. On one hand, several efforts have been made to utilize these materials either as monolithic devices or as a part of a larger structure. Examples include guided mode resonances in atomically thin semiconductors[1], lasers with ultrasmall active regions[2,3], ultrahigh precision transition metal dichalcogenide (TMD) metasurfaces[4] and atomically thin flat lenses[5]. On the other hand, photonic cavities have been used to increase the interaction of matter modes with light. If the interaction strength is strong enough, light-matter hybrid modes (polaritons) can be formed, which have been highlighted by numerous fundamental studies of the light-matter coupling and their potential applications[6–11]. The appearance of polaritons in solids have been reported both with external cavities (Fabry-Pérot[12,13], photonic crystal modes[7,14]) and with self-hybridized cavities[15–17] where the material itself provides the photon confinement in the structure. Cavity engineering has been shown to modify macroscopic properties in equilibrium via feedback on the effective temperature of the system[18] (thermal Purcell effect) and is predicted to affect the ground and excited states of many other quantum materials[19,20].

A promising but less explored aspect of quantum materials is that they can often be controlled on demand using external fields[21–23]. Photonic devices stand to benefit greatly from the incorporation of quantum materials, as the fabrication of photonic nanostructures generally does not allow for in situ tunability of the resulting geometry once the design is set. Of great potential among quantum material-based photonic structures are monolithic devices where modes are able to fully reside in the material. This could translate the unique properties and tunability of quantum materials into the realm of nanophotonics while taking advantage of the maximum overlap of the photonic and matter modes for the strongest light-matter coupling. This is favorable over most common strong coupling examples with proximitized cavities where the coupling strength is necessarily reduced due to the coupling occurring through evanescent fields only[16]. The realization of such monolithic devices is rare, likely due to a combination of insufficient characterization needed for device design, and lack of established protocols for the nanofabrication.

In this work, we design and experimentally demonstrate in situ tunable nanophotonic modes in photonic crystal slabs based on the low-dimensional magnet, CrSBr. We devise a special fabrication method to nanopattern CrSBr flakes and observe the formation of multiple high-Q guided mode resonances and epsilon-below-zero modes that occupy small mode volumes. By applying external magnetic fields, we show how the mode of operation of the photonic crystal can switch from a typical elliptic mode to a hyperbolic one without changing the device geometry, driven by a massive change in the real permittivity ($\Delta\epsilon_1 \sim 800$ near the excitonic resonance).

This work makes two key contributions. We first show that the combination of nanophotonics and novel quantum materials like CrSBr leads to enhanced optical properties, particularly in terms of tunability and drastic miniaturization in a standalone, monolithic device. Second, we show that the photonic modes of the device self-hybridize with the matter modes, altering the optical properties of the sample through polariton formation. Both through the initial nanostructure design and through the in situ application of magnetic fields post fabrication, we observe tunable intrinsic strong coupling over a wide range of parameters including sample thicknesses, patterning design features, and spectral ranges. Thus, we demonstrate a new tuning knob for nanophotonics and cavity engineering in the form of external magnetic fields using monolithic optical devices made of quantum materials.

**Results**

CrSBr is an air-stable van der Waals antiferromagnet that recently sparked interest due to its relatively high Néel temperature ($T_N$ = 132 K) and the large anisotropy in its optical, magnetic, and structural properties[24–27]. We first investigate the magnetic field dependence of the complex permittivity tensor in unpatterned CrSBr flakes. These flakes, exfoliated from bulk crystals onto sapphire substrates with thicknesses ranging from 10 to 300 nm, were systematically analyzed using reflectivity spectra measurements with incident light polarized along the two principal crystallographic directions, *a* and *b*[28]. The real part of the permittivity along the *b*-axis is presented in Figure 1b. We observe two main contributions: one excitonic resonance at around 910 nm, the other at 710 nm, which are labelled as $X_1$ and $X_2$, respectively. The optical properties of $X_1$ are well studied[22,27]. Recent work showed that $X_2$ is characterized by a similar oscillator strength and a more pronounced magnetic field dispersion, but also a reduced lifetime[28–30].

CrSBr flakes are exfoliated onto Si/SiO$_2$ (285 nm) substrates and subsequently patterned by reactive ion etching through silicon nitride masks to obtain photonic crystal slabs. These masks were previously defined by focused ion beam lithography and then transferred onto the CrSBr flakes using polymer-based dry-transfer techniques (Methods, Supp. Fig. 1)[31,32]. This fabrication method does not involve a liftoff step; as a result, the flakes that are weakly bound to the substrate are not damaged or modified post fabrication. To retain the freedom to define an additional breaking of in-plane rotational symmetries in the fabrication design, we etch a periodic array of elliptical holes that can be rotated about an out-of-plane axis.

We measure reflectivity (Fig. 1c) and photoluminescence (PL) (Supp. Fig. 2) from a 30-nm-thick patterned sample (see Fig. 1 for patterning parameters) at 5 K. Optimizing the parameters of the design geometry, we observe several guided mode resonances (GMR), brightened by the periodic patterning, together with a Fabry-Pérot mode solely determined by the flake thickness. The permittivity tensor extracted from the experimental data and prior work are then used as input to RCWA and finite element method solvers for reflectivity simulations. This accurately reproduces the observed spectral response as shown in Fig. 1c (for more details, see Methods). The field profiles of these modes are calculated by finite element simulations and presented in Fig. 1d. As the real part of the refractive index in the 910-930 nm region is extremely large ($n$ = 6 to 20), higher order modes are uniquely very close to one another spectrally (2-3 GMRs in $\Delta\lambda = 20$ nm for these patterning parameters). This contrasts sharply with silicon, where the same patterning would result in the order of hundreds of nanometers of spectral separation. We note that due to the orders-of-magnitude difference in the in-plane and out-of-plane permittivities (Methods), the modes are strongly transverse-electric (TE). As a result, we will refer to them as $TE_{ml}$; see Methods for the convention we use. The $TE_{41}$ resonance has a Q-factor greater than 500, while the $TE_{21}$ resonance Q-factor is around 150. As a result of the anomalously large refractive index near the excitonic resonance, all photonic bands we report for the 30-nm-thick slab are essentially flat in momentum space (Supp. Fig. 2). By the same token, the large refractive index is also responsible for the miniaturization of the device: the waveguide thickness is $\lambda/30$ from 30-nm-thick slabs, which can be enhanced as much as to $\lambda/150$ from GMRs of 6 nm thick CrSBr on SiO$_2$ (Supp. Fig. 3).

The introduced patterns also act like monolithic photonic cavities, and result in self-hybridization between the photonic mode (GMR) and the exciton ($X_1$). The resulting exciton-polaritons are bright in PL and their energy can be tuned by external magnetic fields (Supp. Fig. 2). These patterns allow us to explore light-matter coupling across a wide spectral region, without limitations imposed by slab thickness for Fabry-Pérot (FP) mode formation[17,33]. The Rabi splitting due to $TE_{21}$ in the 30-nm thick sample is $\hbar\Omega_R = 66\ meV$, an order of magnitude larger than $\hbar\Omega_R < 7\ meV$ from the strong coupling of the FP mode to the exciton (Supp. Fig. 2).

We also note that for a large subset of patterns and thicknesses, we observe reflectivity dips in the epsilon-below-zero region ~860-908 nm (see Fig. 1b), indicating the creation of absorption modes. The imprinted patterns generate bright photoluminescence (PL) modes, implying the formation of hyperbolic exciton-polaritons (HEP)[34] which are also tunable with magnetic fields (Supp. Figs. 4-5, 16, also see Supplementary Note 1).

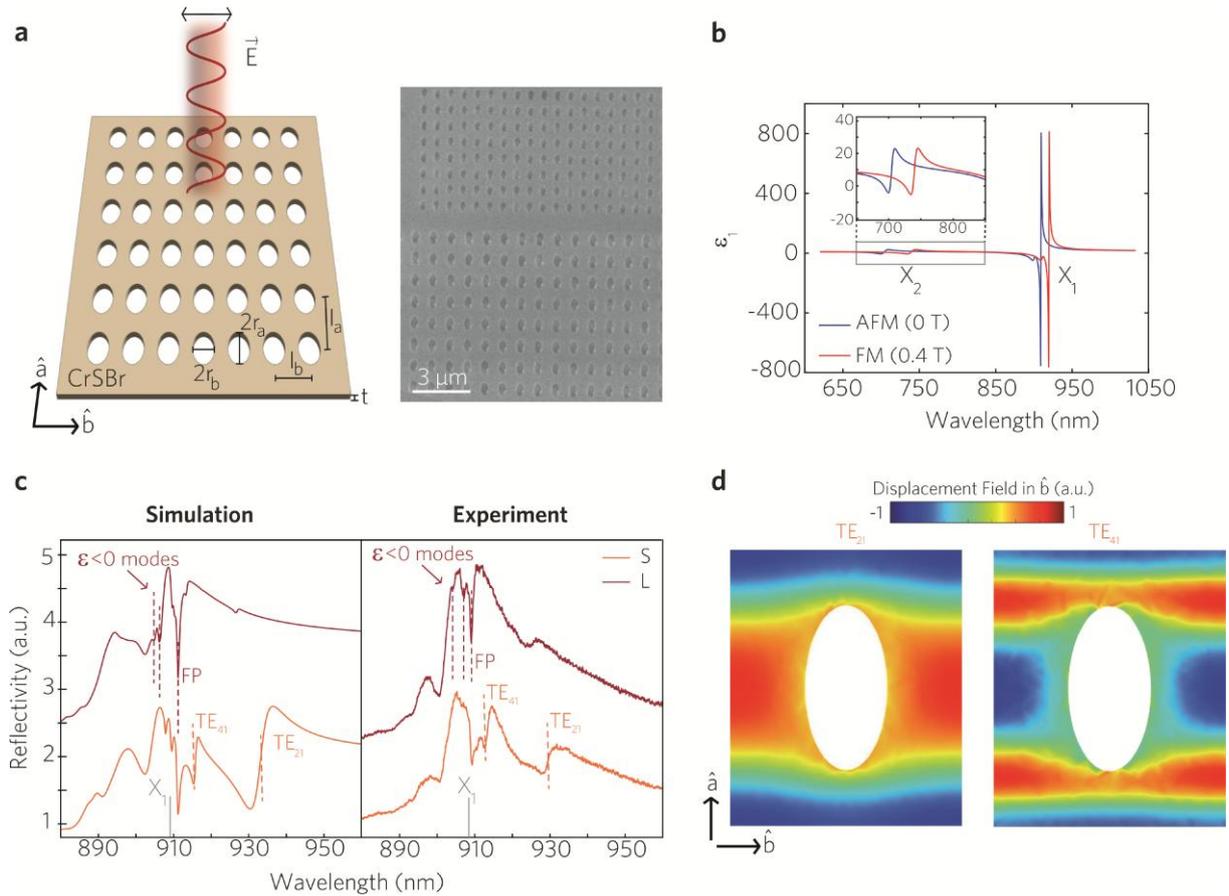

**Figure 1: Photonic resonances in PhC slabs made of CrSBr. a,** Illustration of the patterned CrSBr sample with a representative SEM image (top pattern is L as given below). The light polarization is parallel to the *b*-axis throughout the work. **b,** The permittivity of the *b*-axis of unpatterned CrSBr at 5K, extracted from thickness dependent normal reflectivity. $X_1$ and $X_2$ are the two excitons defined in text. Polarizing the spins of the material by applying a magnetic field of 0.4 T along the *b*-axis switches the ground state from antiferromagnetic (AFM) to ferromagnetic (FM), modifying the permittivity substantially. **c,** Patterning the sample results in brightening the dark modes of the system, allowing the detection of the modes other than FP dips in reflectance. For the larger pattern, "L", (red curve) ($r_a$ = 180 nm, $r_b$ = 90 nm, $l_a$ = 600 nm, $l_b$ = 500 nm, t = 30 nm), discernible modes appear primarily in the epsilon-below-zero region. For the smaller pattern, "S", (orange curve) ($r_a$ = 120 nm, $r_b$ = 60 nm, $l_a$ = 400 nm, $l_b$ = 333 nm, t = 30 nm), a number of TE-like modes appear, in good agreement with the RCWA simulations. **d,** Field profiles of the guided mode resonances of "S" sample in **c** from FEM simulations.

We demonstrate the creation of GMRs in another, thinner (t = 20 nm) sample (Fig. 2). Patterns, with size progression, are transferred onto a large, uniform flake. For this experiment, the ellipses are patterned with a 20 degrees rotation around the out-of-plane axis as shown in the SEM image of Fig 2a.

We again collect reflectance and PL and observe the spectral progression of the coupled waveguide mode with respect to the varying lattice parameters. The patterning weakens the FP mode in the highest index region and brightens the waveguide modes. All features in the epsilon-positive region in the spectrum can be ascribed to GMRs, according to the simulations. The dips observed for the $r_a$ = 90 nm and $r_a$ = 60 nm patterns are wider than the ones observed in Fig. 1, as there are two GMRs that are intriguingly just 3 nm apart due to the real permittivity jumping from 100 to 450 when going from 915 nm to 910 nm in wavelength. This remarkable increase in permittivity pushes the waveguide modes closer to each other. Distinguishing these two modes is challenging even with normal reflectivity simulations (Fig. 2c), which agree well with the experiments. We identify these modes as $TE_{41}$ and $TE_{21}$ (see also Methods, the mode profiles are displayed in Supp. Fig. 6). We also note that the PL maximum (Fig. 2b) follows the reflectivity dip in Fig. 2c, indicating the strong coupling of the exciton $X_1$ to the GMR cavity modes.

We similarly observe PL emission from the epsilon-below-zero region (below 908 nm) that gets more pronounced with the increasing absorption as shown in the reflectivity plot, Fig. 2b. We ascribe this absorption edge, again, to HEPs[34] that appear due to patterning.

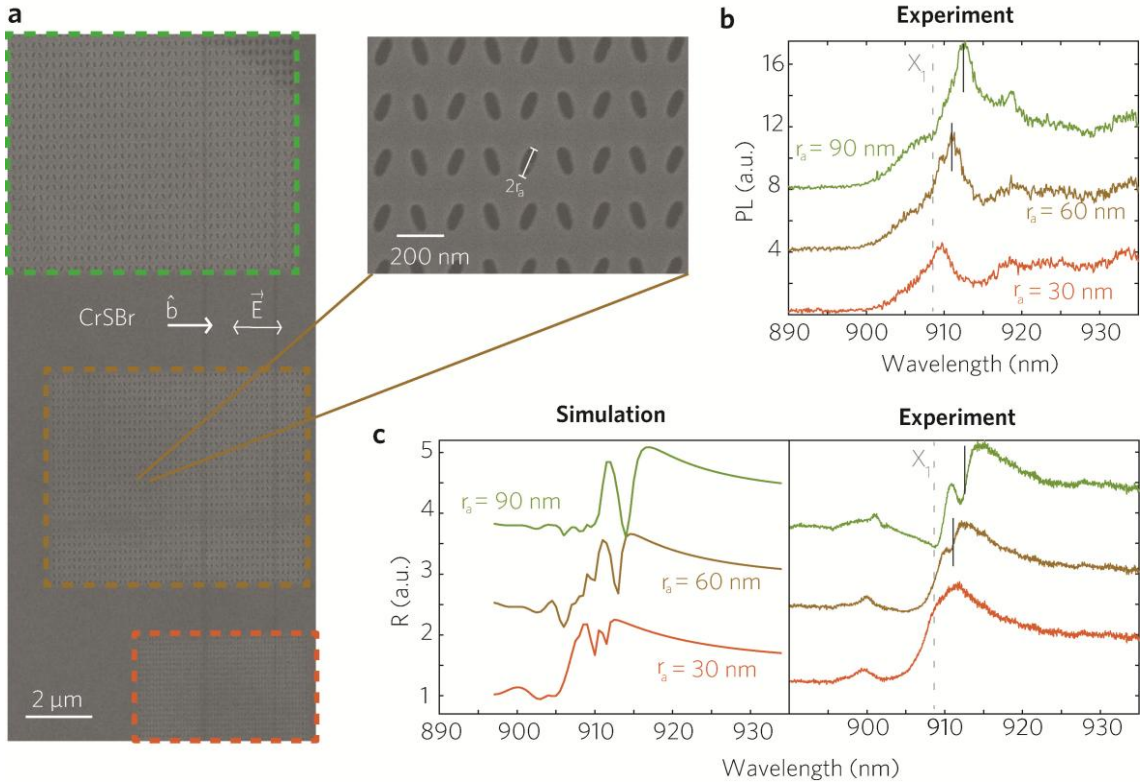

**Figure 2: Self-hybridized exciton-polaritons in CrSBr photonic crystal slabs. a,** SEM image of the sample. Ellipses are rotated by 20 degrees for this sample, as shown. Color code is defined in the next panel. **b,** Photoluminescence from three samples with different patterning parameters. Green: $r_a$ = 90 nm, $r_b$ = 40 nm, $l_a$ = 300 nm, $l_b$ = 200 nm, t = 20 nm; Brown: $r_a$ = 60 nm, $r_b$ = 26 nm, $l_a$ = 232 nm, $l_b$ = 157 nm, t = 20 nm; Orange: $r_a$ = 30 nm, $r_b$ = 14 nm, $l_a$ = 150 nm, $l_b$ = 100 nm, t = 20 nm. The lower energy part of the PL spectrum has a fine structure likely due to exciton-phonon interactions and is out of scope of this study. We note the finite PL from the epsilon-below-zero region. **c,** Reflectivity spectrum of the samples introduced in **b**. Guided resonance maxima follow the PL maxima for all samples, indicating strong coupling of the photonic mode to the exciton. Straight line indicates the guided mode resonances $TE_{21}$ and $TE_{41}$, dashed line indicates the center of $X_1$. We also note the developing dip around 905 nm, indicating allowed absorption modes in the epsilon-below-zero region.

The unique properties of CrSBr arise from both its exceptionally large and anisotropic refractive index and the dependence of these optical properties on its ground state. As shown in Fig. 1b, the 910 nm $X_1$ exciton absorption center redshifts by as much as 10 nm (15 meV) when the magnetic ground state of the material is switched from antiferromagnetic (AFM) to ferromagnetic (FM) when the spins are polarized[22]. Remarkably, the magnetic field required to polarize the spins is also very anisotropic. If the magnetic field is applied parallel to the *b*-axis (the easy axis) of the material, the spins flip abruptly at around 0.34 T. For the *a* and *c* axes, the spins rearrange continuously, and the saturation occurs at around 1 T and 2.4 T, respectively.

We leverage the sensitivity of CrSBr to its environment to tune the photonic resonances we reported above in this work in a reversible manner. We apply a relatively small magnetic field of 0.4 T parallel to the *b*-axis of CrSBr sample shown in Fig. 2. The AFM-to-FM spin flip transition is accompanied by the rigid redshift of the photonic resonances by as much as $\Delta\lambda = 10$ nm. This implies that upon the application of the magnetic field, modes residing around $\lambda = 914$ nm switch from GMRs in the positive-permittivity region to hyperbolic exciton-polariton resonances (Fig. 3a). To our knowledge, this is the first instance of achieving on-demand operational mode switching in a photonic crystal without altering its geometry.

When the magnetic field is turned off, the resonances return to their original position. It is also possible to tune the spectral position of the resonances continuously, by applying a magnetic field in any perpendicular direction (*a* or *c* axes). We exemplify this in Supp. Fig. 4 by shifting the PL emission from the HEPs continuously through the application of magnetic fields parallel to *a*-axis of the crystal. As the origin of the transition is electronic[22], the mode shifting is robust without any limitations on the number of switching operations.

The *b*-axis permittivity of CrSBr also depends on the temperature. As temperature increases from 5K to 55K, the exciton absorption peak redshifts and broadens. This redshift and broadening also affect the resonances due to a weakening of the permittivity contrast (Fig. 3b).

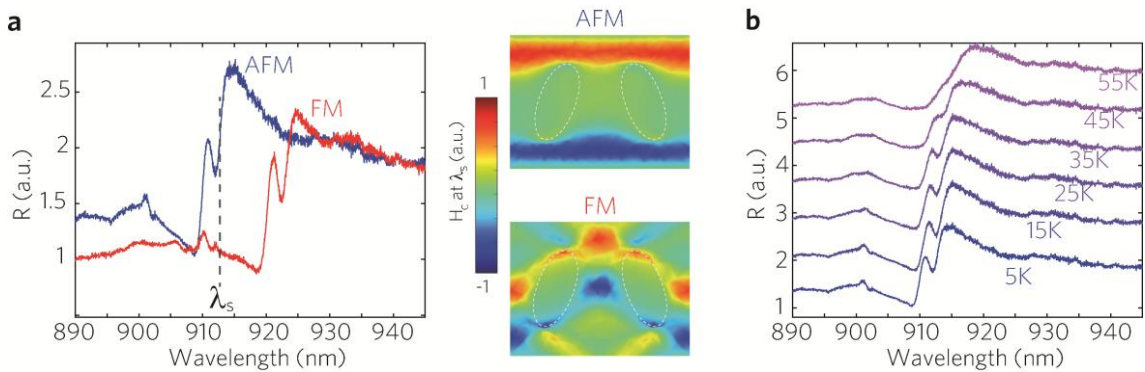

**Figure 3: In situ tunability of the photonic modes. a,** When CrSBr is magnetically polarized, the exciton energy, hence the refractive index, redshifts. The photonic modes redshift accordingly. Here, the GMR of Fig. 2 shifts by 10 nm (15 meV) when a magnetic field of 0.4T is applied along the *b*-axis. The mode switching is illustrated by the mode profiles at a given wavelength of operation. See Supplementary Note 1 for a discussion of the mode symmetries and Supp. Fig. 6 for in-plane electric field mode profiles of the brightened HEPs. **b,** The permittivity depends on the temperature. Increasing the temperature results in a less pronounced permittivity anisotropy, weakening the spectral features.

We illustrate an even stronger tuning of these photonic modes with magnetic fields by utilizing the higher energy exciton, $X_2$. This part of the optical spectrum of CrSBr is less well studied compared to $X_1$, likely due to a combination of weaker PL emission and higher losses around the center of exciton absorption. Nonetheless, the magnetic field dispersion of $X_2$ is almost an order of magnitude stronger than $X_1$ in terms of absolute energy shift, as shown in the permittivity data (Fig. 1b) and in a reference reflectivity measurement on an unpatterned flake (Fig. 4a). This characteristic makes the $X_2$ resonance more interesting for in situ tunability purposes and to form GMRs in the near-visible range.

We perform reflectivity and PL measurements on a patterned slab of CrSBr with thickness 30 nm (Fig 4a-b, Supp. Fig. 8). Similar to GMRs near $X_1$, we observe a reflectivity dip in the high permittivity region (Fig. 4b). Upon application of the magnetic field, the spins are polarized along the $b$-axis while the GMR mode redshifts as much as 25 nm (50 meV), or three times more than $X_1$.

It should be noted that the reflectivity dip is much wider than the $X_1$ resonance. This is expected, as the permittivity is an order of magnitude lower compared to the region of interest around $X_1$. This results in photonic bands being relatively more dispersive in momentum space. Even without momentum dispersion and with normal incidence reflection, the modes exhibit significantly lower Q-factors (Fig. 4c) than near-$X_1$ GMRs, due to increased losses and reduced index contrast.

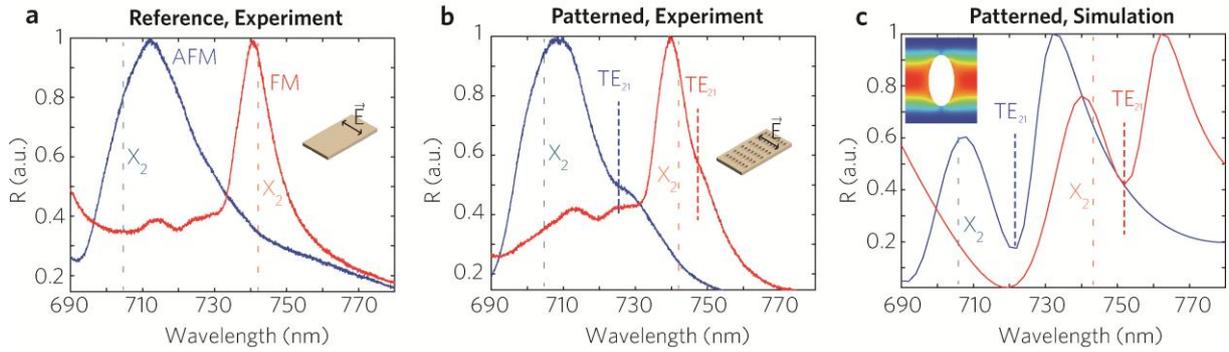

**Figure 4: Strongly tunable photonic resonances in the near-visible range. a,** Reference reflectivity spectra from unpatterned CrSBr, t = 30 nm. The permittivity peak shifts by 30 nm (70 meV) due to the shift of $X_2$ in a magnetic field. Approximate spectral positions of $X_2$ for AFM and FM states are indicated by faint dashed lines. **b,** Patterned samples host GMRs that are magnetic field tunable by 25 nm (50 meV). Patterning parameters: $r_a$ = 120 nm, $r_b$ = 60 nm, $l_a$ = 400 nm, $l_b$ = 333 nm, t = 30 nm. **c,** Normal reflectance simulations of the patterned sample agree with the observed modes. Inset: Displacement field profile in the $b$-axis for the GMR mode ($TE_{21}$ in the AFM phase).

**Discussion**

Altogether, we demonstrate ultrathin (as thin as $\lambda/150$) CrSBr photonic crystal slabs supporting in situ magnetically tunable resonant modes. These guided modes self-hybridize with excitons, forming exciton-polaritons controllable by both structural design and magnetic fields. The findings of this study may be advanced through modifications to the design of photonic crystals or by utilizing different materials in their fabrication. Additionally, the potential for spectral tuning can be harnessed across a significantly broader spectral range.

We opted to work mainly within the thickness range 6-30 nm for the PhC slabs. However, thinner PhC slabs, down to the monolayer limit, are feasible, although perhaps with a lower permittivity environment.

These PhC slabs offer a direct pathway to higher Q-factors through BICs, enabling stronger light-matter interactions and lasing[35–39].

The focus of this work is monolithic optical nanostructures made of CrSBr operating at cryogenic temperatures, which are necessary because the magnetic order sets in only below 132 K. Nevertheless, we stress that the design principles we put forward in this work are not specific to CrSBr, and that materials similar in structure (CrOCl[40], NbOCl$_2$[41,42], CrPS$_4$[43], …) or in optical properties in general may also be utilized, even at room temperature, in nanophotonic structures similar to those investigated here. We note that the optical properties of CrSBr photonic structures can also be tuned by gating[44].

Although the redshift of the exciton wavelengths across the magnetic transition is relatively small (10 nm and 40 nm for $X_1$ and $X_2$ respectively), due to the extremely large oscillator strength of the excitons, the effect of the magnetic fields is not limited to the redshift region. The permittivity contrast due to the transition from AFM to FM (defined as the absolute value of $\Delta\epsilon_B = \epsilon_{1,AFM} - \epsilon_{1,FM}$) is larger than the permittivity contrast between silicon nitride, a widely used semiconductor in photonics, and vacuum in a wavelength range of over 180 nm across the visible and near-IR spectrum (Supp. Fig. 7). A direct implication is the use of CrSBr slabs, even without the strict need for patterning, as magnetic-field activated switches to alter the dielectric medium for other optical components. Through periodic patterning, the effective permittivity of the slabs is modified strongly due to the waveguide modes being folded onto the light cone. This mechanism enables the enhancement or suppression of the permittivity contrast $\Delta\epsilon_B$ where the guided modes appear, facilitating precise spectral tunability in the far-field. This feature makes the PhC slabs discussed in this work ideal for tunable optical filters and modulators that can also be extremely thin and operate at a wide angle (Supp. Fig. 2), due to the unique combination of high refractive index and tunability.

**Methods**

*CrSBr Crystals:* All crystals used in this study are purchased from HQ Graphene. The flakes are obtained by mechanical exfoliation from bulk single crsytals onto Si/SiO$_2$ or sapphire substrates. The approximate thickness of unpatterned flakes can be determined by a combination of apparent color and reflectivity (tracking the number and positions of Fabry-Pérot modes according to the permittivity fits, see *Permittivity Fits*). For more precise thickness information, we use atomic force microscopy (AFM).

*Nanofabrication:* Suspended silicon nitride membranes are purchased from NORCADA (TA301X, 50 nm thick). We used the Raith VELION FIB-SEM, at Characterization.nano of MIT.nano, for the lithographic patterning of the membranes. After inspecting the patterns (with SEM or optical micrography), we transfer these membranes on the sample of choice using well-established polymer-stamp transfer techniques. One technique we find particularly simple is attaching a small piece of polydimethylsiloxane (PDMS) on a glass stamp and picking up the membranes directly from the silicon support. The pick-up success is significantly higher when the membranes are first annealed at 150 degrees Celsius for 10-30 minutes and when PDMS is lifted off very fast after the contact with the cooled-down membrane. Another method we employed is using FIB to mill the edges of the membrane on a flat substrate, and then pick up the membrane using the transfer technique described above. After the transfer, we perform reactive ion etching (Samco Plasma Etching System RIE-230iP) to etch away the parts under the patterns in SiN membranes. We use a chlorine recipe, normally used to etch chromium, to etch CrSBr: Cl$_2$ 20 sccm and O$_2$ 3 sccm. This recipe results in fast etching rates (up to 10 nm/s) for CrSBr, while the SiN is not etched. See Supp. Fig. 1 for an illustration of the fabrication steps. Another, more general, method would be to use argon milling. We confirm that CrSBr is completely etched first by optical images (CrSBr flakes under SiN look the same color while the regions unprotected disappear) and SEM later, where the lack of CrSBr results in a larger contrast especially for thicker samples. We note that we do not remove the SiN membranes after the fabrication; they are incorporated into the structure. As SiN has a very small relative permittivity compared to CrSBr in the spectral regions we are reporting (4 vs 100-450 for X$_1$, 4 vs ~20 for X$_2$), leaving the membrane on CrSBr after the patterning is inconsequential.

*Optical Spectroscopy:* We use a supercontinuum light source (NKT Photonics SuperK Fianum) as the excitation source for reflectivity measurements. Broadband light is first polarized by a polarizer and an achromatic half-wave plate to the *b*-axis of the crystal and then focused on the sample by a 20x objective (NA = 0.25) in a cryostat (Montana S50) equipped with a 1T in-plane magnet. All measurements are performed at 5K unless otherwise noted. Using an objective without back-focal plane imaging would normally be detrimental to the observation of dispersive features in the spectrum because of the momentum spread. However, in this case, the in-plane refractive index is so large that the photonic bands arising from coupling with the excitons are flat, essentially eliminating the need for angle-resolved spectroscopy (see Supp. Fig. 5). The collected light from the sample is then sent to a spectrometer (Horiba LabRAM Evolution). Photoluminescence measurements use the same setup with a 532 nm CW laser instead of the supercontinuum laser. All reflectivity spectra presented in this work are normalized by the broadband reflectance from the 50-nm-SiN/285-nm-SiO$_2$/Si interface.

*Permittivity Fits:* We performed reflectivity measurements on a number of unpatterned CrSBr flakes with varying thickness on sapphire[28]. Due to the extremely large refractive index around X$_1$ and X$_2$, we start observing sequential Fabry-Pérot modes with increasing thickness. We fit a Lorentzian for all the features visible in the reflectivity spectrum for the *b*-axis:

$$\epsilon(E) = \epsilon_b + \sum_i \frac{f_i}{E_i^2 - E^2 + iE\Gamma_i}$$

where $E$ is energy, $\epsilon_b$ is the background permittivity (which we fit as 11.1, similar to refs[33,34]), $f_i$ is the oscillator strength, $E_i$ is the energy of the exciton, and $\Gamma_i$ is damping for the oscillator. We feed these variables into a transfer matrix method (TMM) solver as the permittivity of a slab with a given thickness on sapphire substrate. We measure 16 thin flakes with different thicknesses, ranging from 13 nm to 208 nm, and fit the reflectivity data from TMM to the experimental data. This is relatively straightforward, as even the flakes as thin as 60 nm exhibit multiple Fabry-Pérot dips. See reference[28] for the reflectivity data (with only two oscillators fit). For this work, we also fit a third oscillator, X*, as we are also interested in the hyperbolic permittivity region. The permittivity values given in Fig. 1b are obtained from the following parameters.

$X_1$: $f_{X_1} = 1.7\ (eV)^2, E_{1,AFM} = 1.3637\ eV, E_{1,FM} = 1.3481\ eV, \Gamma_1 = 0.0008\ eV$.

$X_2$: $f_{X_2} = 1.2\ (eV)^2, E_{2,AFM} = 1.76\ eV, E_{2,FM} = 1.67\ eV, \Gamma_2 = 0.025\ eV$.

$X^*$: $f_{X_*} = 0.5(eV)^2, E_{X_*,AFM} = 1.3771\ eV, E_{X_*,FM} = 1.361\ eV, \Gamma_{X_*} = 0.0075\ eV$.

We take the *a*-axis and *c*-axis relative permittivity values as 11 and 4 from prior studies[33,34]. These permittivity values are then used for both FEM and RCWA simulations.

*Simulations:* We use S4[45] for RCWA (Fig. 1c) and COMSOL Multiphysics for FEM simulations (Fig. 1d, Fig. 2c, Fig. 4d), using the permittivity values given above, and the geometry described in the main text: Air/Patterned 50-nm-thick SiN/Patterned CrSBr with a given thickness/285-nm-thick $SiO_2$/Si. All simulations presented in the main text are normal incidence reflectance; we present angular dependence in the Supplementary Materials. $TE_{ml}$ is defined as transverse electric modes that have m antinodes in the electric field profile in the *a*-axis and l antinodes in the *b*-axis. We deduce this from the mode profiles FEM yields. We again note that we call TE-like modes TE in the main text. In a photonic crystal slab, there are no pure TE/TM modes.

**Acknowledgements:** The authors acknowledge helpful discussions with Mehmet Butun. This material is based upon work supported by the U.S. Department of Energy, Office of Science National Quantum Information Science Research Center's Co-design Center for Quantum Advantage (C2QA) under contract number DE-SC0012704. C2QA led in this research. This material is also based upon work sponsored in part by the U.S. Army DEVCOM ARL Army Research Office through the MIT Institute for Soldier Nanotechnologies under Cooperative Agreement number W911NF-23-2-0121. This work was performed in part on the Raith VELION FIB-SEM in the MIT.nano Characterization Facilities (Award: DMR-2117609) and in Fab.nano and Characterization.nano facilities at MIT.nano. A.K.D. acknowledges support from MathWorks Science Fellowship.

**Author Contributions:** A.K.D., L.N., and R.C. conceived the project. A.K.D. performed the nanofabrication and carried out the experiments together with L.N., with help from C.A.O., under the supervision of R.C. S.V., supervised by M. S., and A.K.D. performed the simulations. All authors contributed to the writing of the manuscript.